\documentclass[final,english]{bullsrsl}[2022/06/15]

\usepackage[utf8]{inputenc}
\usepackage[T1]{fontenc}

\usepackage[numbers]{natbib} 
\usepackage{graphicx}

\begin{document}
\title{The cradle of nonlinear asteroseismology: observations of oscillation mode variability in compact pulsating stars}

\author[affil={1,2},corresponding]{Weikai}{Zong}
\author[affil={3},corresponding]{St{\'e}phane}{Charpinet}
\author[affil={3}]{G{\'e}rard}{Vauclair}
\author[affil={1,2}]{Jian-Ning}{Fu}
\author[affil={1,2}]{Xiao-Yu}{Ma}
\affiliation[1]{Institute for Frontiers in Astronomy and Astrophysics, Beijing Normal University, Beijing 102206, China}
\affiliation[2]{Department of Astronomy, Beijing Normal University, Beijing 100875, China}
\affiliation[3]{IRAP,~CNRS,~Universit\'{e} de Toulouse, CNES,~14 Avenue Edouard Belin,~31400 Toulouse,~France}
\correspondance{weikai.zong@bnu.edu.cn; stephane.charpinet@irap.omp.eu}
\date{13th October 2020}
\maketitle


%

\begin{abstract}
We briefly review progress in developing a pathway to nonlinear astereoseismology, both from theoretical and observational aspects. As predicted by the theory of weak nonlinear interactions between resonant modes, their amplitude and frequency can be modulated according to various kinds of patterns. However, those subtle modulations could hardly be well characterized from ground-based photometric monitoring. The {\sl Kepler} spacecraft offered a new window to find clear-cut evidence of well-determined amplitude and frequency modulations, leading to the first discoveries of such variations in pulsating white dwarf and hot B subdwarf stars. Following that direction, a systematic survey of oscillation mode properties in compact pulsators monitored by {\sl Kepler} suggests that mode variability is likely a common phenomenon, which remain unaccounted for by standard linear non-radial pulsation theory. To reach this conclusion firmly, the survey has now been extended to a larger context including compact stars observed by K2 and TESS. We expect that this extended survey will help to constrain key parameters governing weak nonlinear effects in stellar oscillations.
\end{abstract}


\msccodes{---- ---- ----}





\section{Past}
The first developments in nonlinear asteroseismology can be traced back before the 1980s, when nonlinear approaches were natural to consider higher-order terms in perturbation calculations of stellar hydrodynamics. Among various treatments for nonlinear terms, the formalism of amplitude equations (AEs) was solvable and attractive, as it reduces the problem of coupled partial differential equations to ordinary differential equations only in time. Around the 1980s, both non-adiabatic and adiabatic formulations of AEs were developed, but applied to simple cases : for instance, three resonant mode interactions \cite{1982AcA....32..147D} or radial mode pulsations \cite{1984ApJ...279..394B}. Even though these early calculations of AEs were relatively simple, they provided new insight on the dynamical behavior of interacting pulsation modes. The parametric resonance, involving parent driven modes and daughter damped modes, can reproduce temporal limit cycles of amplitude and frequency variations under certain conditions \cite{1985AcA....35..229M}. The non-adiabatic radial solutions of AEs agree quite well with the models constructed for classical Cepheids \citep{1986ApJ...303..749B,1989ApJ...346..898K}.

In the 1990s, non-adiabatic AEs were applied to non-radial resonant modes using both Eulerian and Lagrangian forms independently \cite{1994A&A...291..481G,1994A&A...292..471V}. Numerical exploration of three-mode interactions were successfully conducted for the case of triplets produced by rotation \cite{1995A&A...296..405B,1997A&A...321..159B}, predicting that the modes involved exhibit three major distinct behaviors depending on their nonlinear coupling parameters. A specific calculation of parametric AEs was applied to pulsating white dwarf stars, with predictions concerning the dynamic and amplitude saturation of the oscillation modes \cite{2001ApJ...546..469W}. First applications of AEs were on pulsating white dwarf and $\delta$~Scuti stars due to their weak non-adiabatic pulsations \cite{1998BaltA...7...21G}. However, nonlinear AEs eventually passed out of attention due the lack of secure observational constraints available at that time. Limited by this, we note that AEs have only been applied to RR~Lyrae stars after the turn of the new century, in an attempt to solve the Blazhko-effect problem \cite{2011ApJ...731...24B}. 

As a matter of fact, nonlinear effects on oscillation modes can hardly be captured by ground-based monitoring for two main reasons: the induced variations occur on long timescales and the weak modulation amplitudes are not easily resolved correctly. Despite these difficulties, the quest to link observed behaviors with predictions from nonlinear seismic calculations started at the same epoch as theoretical works were developed \cite{1989A&A...215L..17V}. It was recognized that pulsating white dwarfs could be among best candidates to test nonlinear AEs, as hints of amplitude changes and variations in rotational splittings, for instance in GD~358, were identified \cite{1998BaltA...7...21G}. Direct observation of nonlinear resonant mode couplings may have been found in a hot white dwarf, PG~0122+200, in which mode amplitude and frequency variations on much shorter timescales than expected from secular evolution were measured \cite{2011A&A...528A...5V}. However, to really overcome the main obstacles in this field, continuous high-precision photometry covering several years was needed. This was offered by the successful launch of the {\sl Kepler} space telescope \cite{2010Sci...327..977B}.

\section{Present}
When analyzing pulsating hot B subdwarf (sdB) stars observed by {\sl Kepler} using sliding Fourier transforms (sFT \cite{2012A&A...544A...1T}), or equivalently sliding Lomb-Scargle Periodograms (sLSP) implemented in our code \texttt{Felix} \cite{2010A&A...516L...6C}, we detected clear amplitude variations \cite{2015ASPC..493..261Z}. However, the sLSP (or sFT) only shows a rough overview of the dynamics affecting mode amplitudes. In 2013, a dedicated project was initiated by us to characterize amplitude modulations in compact pulsators using {\sl Kepler} photometry in order to investigate nonlinear effects in stellar oscillations.

Based on a continuous, 23-month duration light curve, we analyzed amplitude modulations (AMs) in the pulsating helium-atmosphere DB white dwarf, KIC~08626021, in which periodic patterns were accurately characterized for the first time \cite{2016A&A...585A..22Z}. Most importantly, with the sharp frequency resolution achieved, several extracted frequencies showed a modulation pattern similar to that of AMs, with almost identical modulating periods of the order of a year, as expected from nonlinear theory of weak resonant interactions. We stress that there were no characterization of frequency modulations (FMs) reported before that work, due to limitations on the precision of frequency measurements before the {\sl Kepler} era. After a series of bootstrap tests on these space data, we established that a secure (yet conservative) threshold for signal detection should be set at the level of 5.6$\sigma$ (instead of 4$\sigma$ routinely used with ground-based photometry; $\sigma$ being the average noise in Fourier space), which leaves eight independent frequencies in KIC~08626021. An additional frequency from the previous seven reported in the literature \cite{2014ApJ...794...39B} was identified with the help of nonlinear AM and FM patterns. This analysis further led to the seismic cartography of KIC~08626021 \cite{2018Natur.554...73G}. However, when we use the seismic model obtained for that star to calculate nonlinear mode couplings, significant differences between theoretical expectations and observation exist, indicating that there is room for further exploration in nonlinear parameter space. 

The second case exposing nonlinear AMs and FMs was the pulsating sdB star KIC~10139564 in which diverse modulation patterns in {\sl p}- and {\sl g}-mode multiplets were found \cite{2016A&A...594A..46Z}. An interesting feature in this star is that several dominating triplets exhibit quasi-periodic variations with the superposition of two different components. This suggests that rotational multiplets not only show the nonlinear rotational $1:1:1$ resonance, but also the direct $\omega_1 \sim \omega_2 + \omega_3$ resonance. Thus far, for simplicity in nonlinear theoretical calculation, a mixture of two different types of resonance has never been considered, because it would introduce many complex terms in the AEs \cite{1995A&A...296..405B}. The case of KIC~10139564 suggests however that observed behaviors are more complicated than these simple theoretical treatments. At this stage, AMs and FMs are well characterized in two different types of pulsating stars. Instead of introducing new physical processes to explain their behaviors, we think that nonlinear AEs may offer the most natural path to account quantitatively for this phenomenon.

\begin{figure}
\centering
\includegraphics[width=\textwidth]{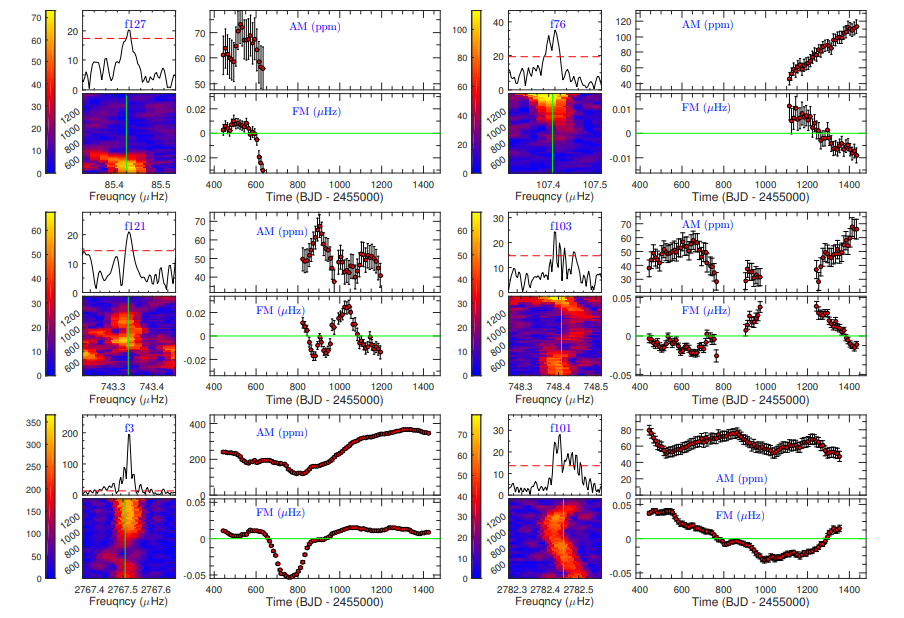}
\caption{\footnotesize{Examples of AMs and FMs for six pulsation modes detected in KIC~3527751. See details in Ref. \cite{2018ApJ...853...98Z}}}
\label{fig}
\end{figure}

Pursuing in this direction, we discovered a variety of AMs and FMs in another sdB star, KIC~3527751, that shows more than one hundred frequencies with high enough amplitudes for such analyses, thus involving more situations than just rotational multiplets \cite{2018ApJ...853...98Z}. The analysis established that all frequencies are not stable during the 3-yr duration of observation. Figure\,\ref{fig} shows six representative examples of AMs and FMs with obvious variations. Some of the frequencies can only be detected in parts of the observing run, either in the beginning, in the middle, or at the end. Interestingly, one can see correlations directly from the AMs and FMs, whose existence was predicted from solutions of nonlinear AEs \cite{1985AcA....35..229M}. Those discoveries pose a new challenge to the physical interpretation of phase modulations (PM) as the signature of stellar or planetary companions around a pulsating star \cite{2007Natur.449..189S,2018A&A...611A..85S}.
 
The results discussed above should be a strong motivation for a renewed attention to nonlinear AEs' calculations. In this context, a thorough comparison of intrinsic AMs and FMs should be evaluated as the {\sl Kepler} pipeline provides different flux calibrations. Detrending processes used to extract the {\sl Kepler} photometry might affect the modulating patterns of AMs and FMs that we seek to measure accurately. In order to test this, we selected two sdB stars, KIC~2438324 and KIC~11179657, as representative benchmark targets \cite{2021ApJ...921...37Z}. Both stars are binaries and we used the orbital signal as a referential for the intrinsic AM and FM comparison. We found that AMs may be different if one chose different photometric flux calibrations, whereas FMs are independent of it and remain the same. In addition, AMs may also suffer from systemic patterns with a timescale identical to {\sl Kepler}'s orbital period. Those findings suggest that nonlinear AEs can more reliably be constrained from the observed FMs and not strictly from the AMs.

Limited by the number of targets in the original {\sl Kepler} sample, we extended investigations to AMs and FMs in white dwarf and sdB stars observed by K2. Compared to {\sl Kepler}, the larger sample of K2 was observed only with a duration of $\sim80$~days, meaning that we can only measure relatively short-term AMs and FMs. A recent work suggests that oscillation modes exhibiting AMs and FMs are indeed very common in the pulsating sdB star EPIC~220422705 \cite{2022ApJ...933..211M}. We fitted the modulating patterns with a few simple profiles for statistical evaluation. Moreover, to remove pollution due to beating of unresolved nearby frequencies from AMs, we introduced a novel approach that subtracts simulations of the modulations due to beating to the observed results. We ultimately found that at least 17 of 22 frequencies have intrinsic AMs in EPIC~220422705, with several occurring over a modulating period of the order of months. These findings provide a very important complement to characterize modulation timescales in compact pulsators. Their values are related to the linear growth rates of pulsation modes, which has never been obtained through observation, yet. At this stage, K2 targets can also be very important to characterize AMs and FMs, which will feed the development of nonlinear AEs in the future.

\section{Future}
Asteroseismology has brought many new insights to better shape the theory of stellar structure and evolution \cite{2021RvMP...93a5001A}, especially thanks to the revolution brought by photometry from space telescopes \cite{2010Sci...327..977B}. However, almost all achievements are triggered by "linear asteroseismology", that is by analyzing and interpreting the frequencies of pulsation modes in the framework of linear pulsation theory. "Nonlinear asteroseismology", that seeks to exploit in addition nonlinear behaviors of modes such as temporal amplitude and frequency variations, has also the potential to provide a variety of information useful to probe stellar interiors. The road map for nonlinear asteroseismology could take several directions, some being covered by our ongoing projects.

Increasing the sample size of compact pulsators is underway as data flood from ongoing TESS mission, which will likely continue with future PLATO and ET2.0 missions \cite{2014ExA....38..249R,2022arXiv220606693G}. Several detailed studies have shown that pulsating white dwarf and sdB stars are arguably the best candidates on that front. A global statistical picture of the phenomenon of AMs and FMs induced by nonlinear effects has to be constructed, possibly leading to evaluations of linear growth rates of oscillation modes in evolved compact stars. However, before such statistics work, we must correctly assess the fraction of pollution affecting intrinsic AMs by exploiting the per-pixel data. A statistical study of FMs will significantly benefit to research fields applying the PM technique, in particular as a way to detect companions and secular evolution effects. Another application could be to help identify modes, as nonlinear resonant modes should experience very similar AM and FM patterns. Such applications could attract attention from various research fields, for instance, planetary science or new physics related to white dwarf cooling. 

As nonlinear AEs are independent of the type of pulsating star, it is natural to expect similar AMs and FMs in different kind of pulsators across the HR diagram. Recently, several works report on the study of nonlinear resonant mode couplings in $\delta$~Scuti stars \cite{2015A&A...579A.133B}, slowly pulsating B stars \cite{2021A&A...655A..59V}, and eclipsing binaries with pulsating components \cite{2020ApJ...896..161G}. These efforts extend significantly potential applications of nonlinear asteroseismology, since thousands of pulsators of all kinds were (and still are) monitored by TESS and {\sl Kepler} missions. This gives opportunities to derive the modulation timescales of AMs and FMs in various physical conditions from a statistical overview based on all kinds of pulsators. The ultimate goal of such projects is to motivate further development of the nonlinear theory of stellar oscillations, and provide new physical constraints to probe stellar interiors. We end this review with the statement that ``the time of renaissance for nonlinear asteroseismology" has come. This declaration might sound bold, but properly reflects the era that this field is now entering in.

\begin{acknowledgments}
We acknowledge the support from the National Natural Science Foundation of China (NSFC) through grants  11903005, 11833002, 12090040 and 12090042. W.Z. is supported by the Fundamental Research Funds for the Central Universities. 
S.C. is supported by the Agence Nationale de la Recherche (ANR, France) under grant ANR-17-CE31-0018, funding the INSIDE project, and financial support from the Centre National d'Études Spatiales (CNES, France).
\end{acknowledgments}

\begin{furtherinformation}

\begin{orcids}
\orcid{0000-0002-7660-9803}{Weikai}{Zong}
\orcid{0000-0002-6018-6180}{St\'ephane}{Charpinet}
\orcid{0000-0001-8241-1740}{Jian-Ning}{Fu}
\end{orcids}

\begin{authorcontributions}
W.Z. leaded and S.C. contributed all projects to this review. S.C. and G.V. supervised W.Z. for his PhD thesis. G.V. and J.F. partly contributed to the projects. X.M. leaded to the last project. All authors commented on the manuscript.
\end{authorcontributions}

\begin{conflictsofinterest}
The authors declare no conflict of interest.
\end{conflictsofinterest}

\end{furtherinformation}

\bibliographystyle{bullsrsl-numen}

\bibliography{bibversion2}
\end{document}